\documentclass[prx,twocolumn,showpacs,amsmath,amssymb]{revtex4-1}

\usepackage{color}
\usepackage{graphicx}
\usepackage{dcolumn}
\usepackage[caption=false]{subfig}
\usepackage{bm}
\usepackage{float}
\renewcommand{\sin}{\mathrm{sin}}
\renewcommand{\cos}{\mathrm{cos}}
\usepackage{hyperref}
\hypersetup{
     colorlinks   = true,
     citecolor    = blue,
     linkcolor    = blue
           }

\preprint{submitted to Physical Review B}

\begin{document}

\title{Parameterisation of non-collinear energy landscapes in itinerant magnets}

\author{A. Jacobsson$^{1}$}\email{adam.jacobsson@ltu.se}
\author{G. Johansson$^{1}$}
\author{O. I. Gorbatov$^{1}$}
\author{M. Le\v{z}ai\'{c}$^2$}
\author{B. Sanyal$^3$}
\author{S. Bl\"{u}gel$^2$}
\author{C. Etz$^1$}

\affiliation{
$^1$Applied physics, Department of Engineering Sciences and Mathematics, Lule\aa~University of Technology, Lule\aa, Sweden\\
$^2$Peter Gr\"{u}nberg Institut and Institute for
Advanced Simulation, Forschungszentrum J\"{u}lich and JARA, 52425
J\"{u}lich, Germany \\
$^3$Department of Physics and
Astronomy, Uppsala University, Box 516, 75120 Uppsala, Sweden 
}

\begin{abstract}
The magnetic force theorem provides convenient ways to study exchange interactions in magnetic systems. However, it is well known that short range interactions in itinerant magnetic systems are poorly described with the conventional use of the theorem and numerous strategies have been developed over the years to overcome this deficiency. In this study, we discuss this issue in the context of the frozen magnon method and find that a self-consistent approach is in general preferable.  
Moreover, an extended Heisenberg model is suggested in order to better describe finite deviations from the magnetic ground state and is shown through cross-validation to give a superior description of the interactions in non-collinear magnetic configurations compared to the regular Heisenberg model. The present study thus supplies a fully self-consistent method for systematic investigations of exchange interactions beyond the standard Heisenberg model. This may prove relevant to high-throughput computational materials science, e.g., in developing high moment materials for the magnetic storage industry.  



\end{abstract}

\maketitle

\section{Introduction}
In the view of solid-state applications, it is of utmost importance to be able to accurately calculate magnetic exchange interactions and predict the correct critical temperatures for all kinds of materials in different geometries (from bulk, to thin films, to nano-structures).  
This is the reason why the present study focuses on developing a method that can properly describe the magnetic interactions in a wide range of systems, with collinear or non-collinear magnetic structures. 

Building on the formulation of Andersen's force theorem\cite{efermi,Heine,KubandMe} the works by Liechtenstein et al.~\cite{Liechtenstein:1984br} and Oswald et al.~\cite{Oswald:1985hs} introduced the magnetic force theorem (MFT). The general formulation by Liechtenstein et al.~\cite{Liechtenstein:1984br,Liechtenstein:1987br} is one of the most widely used methods for the determination of inter-atomic exchange interactions since it may be applied to any collinear magnetic configuration. The great utility of the theorem for the electronic structure community is due to the fact that the exchange interactions are determined from non-self consistent calculations that are orders of magnitude faster than most self-consistent approaches. 

Developments in time-dependent density functional theory (TD-DFT) and many body perturbation theory have made it possible to access the full dynamical magnetic susceptibility~\cite{dyn1,dyn2,dyn3,dyn4}, which gives detailed information on both magnon and Stoner excitations, the latter not being accessible by the different versions of MFT calculations that operate in the adiabatic approximation. However, by a multi-code and multi-scale approach using \textit{ab initio} calculations, Monte Carlo and atomistic spin-dynamics simulations, magnon excitations and magnetic phase transitions may be accurately simulated for larger systems\cite{Spindyn,review,Lezaic:2013ea}.

The MFT relies on the assumption of small changes in the magnetisation and charge density. In addition, usually the internal magnetic fields are varied in the calculations rather than the directions of the moments. Short wavelength excitations are, therefore, determined less accurately and the calculations are said to be performed in the long wavelength approximation (LWA) (for a more extensive discussion see the work of Antropov et al.~\cite{Antropov,antro-shilf}). Various strategies to avoid the LWA have been suggested over the years\cite{antro-shilf,Antropov,Bruno:2003ir,Katsnelson:2004ha}. In particular, developments by Patrick Bruno~\cite{Bruno:2003ir} using "constrained" density functional theory introduced by Dederichs  et al.~\cite{Dederichs:1984bh}, included constraining magnetic fields acting on each atomic site. These fields reinforce the directions of the small transverse spin displacements used to probe the exchange interactions. It was suggested that by neglecting the constraining fields, one obtains a set of "bare" exchange parameters, which differ from the real exchange parameters. The inclusion of the constraining fields leads to a 'renormalisation' of the spin-wave spectrum and the Curie temperature. The work by Katsnelson et al.~\cite{Katsnelson:2004ha} showed that while the renormalisation suggested by Bruno~\cite{Bruno:2003ir} gives better thermodynamic properties, the improvement in the magnon dispersion is questionable since the renormalisation should be small in every case where the adiabatic approximation is valid. Moreover, it is also interesting to compare the results with those obtained using the disordered local moment (DLM) formalism~\cite{gyorffy1985,pindor1983} that also operates beyond the LWA. 

The MFT is usually applied in the frozen magnon method~\cite{Halilov}, where exchange interactions are determined from energy differences or evaluation of the spin-torque for different spin spiral configurations. It has been noted that in the absence of constraining fields (converged in self-consistent spin spiral calculations), a mismatch is introduced between the desired and the resulting directions of the magnetic moments, thus leading to a significant un-physical spin-torque.~\cite{Grotheer:2001db,Grotheer:2000gp} Analogous to Bruno's correction, this can be solved by adding an additional term to the potential, that includes the constraining fields. This procedure will be referred to as the \textit{corrected frozen magnon method} in the continuation of the text. It should be noted that the constraining fields are equal to the Heisenberg exchange fields that act on the magnetic moments. It is, therefore, quite possible to derive the exchange parameters directly from the self-consistently converged constraining fields~\cite{Grotheer:2001db,Grotheer:2000gp}. This approach is used in the present work to formulate the \textit{transverse-field method} that is found to be  preferable to the frozen magnon procedures due to better scaling and the avoidance of a frozen potential.  A possible exception could be for systems with strong relativistic effects, where spin-orbit coupling can be introduced through perturbation theory~\cite{Heide2,Heide}. Furthermore, it is well-suited to parameterise finite deviations from the ground state since no assumption of small changes in magnetisation and charge density is necessary. 

At this point, the next step was to extend the Hamiltonian by including multi-spin interactions and see to what extent the increased flexibility improves the predictive power of the model for finite deviations. Interactions beyond the Heisenberg model have made the focus of many studies in the field of materials science.~\cite{Lounis2010,Szilva2017} However, when constructing effective models from a fit to calculated data it is important to avoid overfitting. A simple but effective strategy to deal with this issue is to employ cross validation~\cite{learn-from-data}, where the calculated data is not only used to parameterise an effective model but also to estimate the predictive power of the model.  

When one aims at an automatic scan of the magnetic behaviour of a large number of compounds, following the lines of high-throughput computational materials design~\cite{Curtarolo2013}, it is advantageous to have a general Hamiltonian and a method of parameterisation that can accurately capture the magnetic phase space for many different compounds at a reasonable computational cost. The transverse field method presented in this work fullfils those requirements. Given a computationally designed, still non-existing material, the developed method allows for a fast calculation of exchange parameters and, importantly, the cross-validation offers a programmable decision maker that will automatically determine which interactions in the extended Heisenberg model are important for the material and calculate them all.

\section{Theory}

\subsection{The frozen magnon methods}

 In the augmented plane-waves (APW) based methods the crystals are divided into spherical muffin-tin regions around the atomic sites and interstitial regions between the sites. 
The constraining fields (${\bf B}^{\text{C}}_{i}$) are introduced in the Kohn-Sham equation, 

\begin{equation}
\left\{-\frac{\hbar^2}{2m}\nabla^2+V_{eff}({\bf r})+{\bf \sigma}\cdot {\bf B}_{eff}({\bf r})-\epsilon_{\nu}\right\}\psi_{\nu}({\bf r})=0
\label{KS}
\end{equation}%
as uniform vector fields in the muffin-tins in addition to the magnetic field ${\bf B}^{XC}({\bf r})$ generated by the exchange-correlation potential. 

\begin{equation}
{\bf B}_{eff}({\bf r})= {\bf B}^{\text{C}}_{i}+{\bf B}^{XC}({\bf r})
\label{EB}
\end{equation}

The constraining fields ${\bf B}^{\text{C}}_{i}$ are the Lagrange multipliers necessary to perform the energy minimisation under the constraint of the specific magnetisation density.~\cite{S-Review} In the self-consistency cycle they are converged together with the rest of the potential such that the components of the total integrated magnetic moments perpendicular to the chosen direction of the moments are zero.

\begin{equation}
{\bf B}^{\text{C}}_{i,p+1} = {\bf B}^{\text{C}}_{i,p} + \lambda ({ {\bf m}_{i,p}} - { {\bf m}_{i,chosen}}) ,
\label{eq:condition}
\end{equation}

where $i$ is the site index, ${\bf m}_i$ is the magnetic moment at site $i$,  $p$ is the iteration number and $\lambda$ is a scaling factor.  The procedure still allows for intra-atomic non-collinearity. 

When calculating the total energy of a cone-spin-spiral, it is usually necessary to add constraining magnetic fields to each magnetic sub-lattice in order to preserve the cone-angle during the self consistency cycles. Exceptions are  systems with the cone-spin-spiral ground state~\cite{Cone} since the constraining fields are equal to the transverse part of the Heisenberg exchange fields. Non-zero transverse exchange fields imply that the system is not in equilibrium.

In the frozen magnon methods, a number of reference $\Gamma$-point spin configurations are converged self-consistently. Configurations where there is a non-zero cone angle of the magnetic moment on either a single magnetic sub-lattice or a pair of magnetic sub-lattices are considered here. These are only collinear if all magnetic sub-lattices are tilted with the same angle. In some previous publications,~\cite{Lezaic:2013ea, Jacobsson:2013dz} these reference states were obtained by rotations of the potential of a collinear state. Since only the potentials in the muffin-tins were rotated, this left a discontinuity in the magnetisation density at the border between the muffin-tins and the interstitial region. It was therefore found that better agreement with self-consistent calculations was achieved by setting the interstitial magnetisation to zero.~\cite{Lezaic:2013ea} 

The procedure followed in this work makes it possible to keep a non-zero interstitial magnetisation since we don't perform any rotations of the potential. On the other hand, it leads to a moderate increase in computational cost for systems with more than one magnetic sub-lattice since several self-consistent non-collinear calculations have to be carried out. Once the potentials for the reference configurations are obtained, non-self consistent total energy calculations are done for a set of spin-spiral wave vectors using the converged $V_{eff}({\bf r})$ and ${\bf B}^{XC}({\bf r})$ from the reference configurations and ${\bf B}^{C}({\bf r})$ set to zero. The total energy differences $\Delta E ({\bf q})$ are approximated as the difference in the sums of eigenvalues through the MFT:~\cite{Liechtenstein:1987br} 

\begin{equation}
\Delta E ({\bf q})\approx \sum_{\nu} \epsilon_{\nu}({\bf q})-\sum_{\nu} \epsilon_{\nu}({\bf 0})
\end{equation}

In the last step, exchange parameters $J_{ij}$ of a Heisenberg Hamiltonian are obtained through a least square fitting procedure:~\cite{Jacobsson:2013dz}

\begin{equation}
\Delta E ({\bf q})=\frac{1}{2} \sum_{i\neq j}J_{ij}({\bf e}_{i}({\bf q})\cdot {\bf e}_{j}({\bf q})-{\bf e}_{i}({\bf 0})\cdot {\bf e}_{j}({\bf 0}))
\label{DeltaE}
\end{equation}

Here the direction of the magnetic moments of a sub-lattice with a cone-angle $\theta$ is given by the expression:

\begin{equation}
\begin{split}
{\bf e}_{i} ({\bf q})& = \sin (\theta_{i}) \cos({\bf q}\cdot {\bf R}_{i} + \phi_{i})\mathrm{\bf x} \\
& ~+ \sin (\theta_{i}) \sin({\bf q}\cdot {\bf R}_{i}+\phi_{i})\mathrm{\bf y} \\
& ~+ \cos(\theta_{i})\mathrm{\bf z} \\
\end{split}
\label{eqn3}
\end{equation}
where $\phi_i$ are the phase factors, {\bf q} is the spin-spiral wave-vector and ${\bf R}_i$ are the lattice vectors.
In the corrected frozen magnon method an extra calculation step is done compared to the conventional method. First, $V_{eff}({\bf r})$ is kept fixed and ${\bf B}^{C}({\bf r})$ is converged separately for each spin-spiral wave vector. In the final total energy calculations, $V_{eff}({\bf r})$ and ${\bf B}^{XC}({\bf r})$ are kept fixed, as in the conventional method, but now the pre-converged ${\bf B}^{\text{C}}_{i}$ are added for each one-shot calculation of the sums of eigenvalues.  There are slight numerical differences between the constraining fields obtained in this way compared to fields obtained fully self-consistently. However, this procedure is much faster than a fully self-consistent calculation and no significant difference was obtained with respect to the total energy differences for the different compounds studied in this work. A noteworthy difference between the two frozen magnon methods is that, while ${\bf e}_{i}$ is roughly parallel to ${\bf B}^{XC}({\bf r})$ in both cases, only in the corrected method is ${\bf e}_{i}$ also guaranteed to be the direction of the integrated moment after a single iteration of the electronic structure code. For materials and magnetic states, where the self-consistently converged magnetic fields fullfil ${\bf B}^{\text{C}}_{i}~\gtrsim~{\bf B}^{XC}({\bf r})$, the correction to the conventional frozen magnon method can be expected to be important. The constraining field ${\bf B}^{C}({\bf r})$ compensates for the spin-torques the system experiences when it is forced to assume magnetic configurations different from the ground state structure. The correction will thus tend to be more relevant for itinerant magnetic systems with small moments, materials with strong exchange interactions or magnetic configurations far from the  ground state.

\subsection{Exchange parameters from a transverse field}

In order to obtain exchange parameters from self-consistent quantities one could perform fully self consistent spin-spiral total energy calculations and extract the exchange parameters from Eq.~\eqref{DeltaE}. This is however computationally expensive even for systems that can be described with a small unit-cell if we want to use an accurate all-electron code. A faster procedure, that is also fully self-consistent, was worked out by Grotheer et al.~\cite{Grotheer:2001db,Grotheer:2000gp}. Their approach exploits the fact that the constraining fields are equal and opposite to the transverse part of the Heisenberg exchange fields. We use this as a starting point and develop the so-called \textit{transverse field method} (TF), in the following. The exchange field ${\bf H}_{0i}$ acting on a magnetic moment ${\bf m}_{i}$ is given by:

\begin{equation}
{\bf H}_{0i}=-\frac{1}{ m_{i}} \sum_{j}J_{ij}{\bf e}_{j}
\label{eq:H0}
\end{equation}

Once the constraining fields and magnetic moments are converged for a number of magnetic configurations, we can solve the following system of equations for the exchange parameters, $J_{ij}$:

\begin{equation}
\begin{split}
{\bf B}^{\text{C}}_{i} \cdot {\bf B}^{\text{C}}_{i} =  \frac{1}{ m_{i}} \sum_{j}J_{ij} {\bf B}^{\text{C}}_{i} \cdot {\bf e}_{j}  \\
\end{split}
\label{eq:susc2}
\end{equation}

Relative to previous studies, besides the difference in starting point (i.e. real space vs. reciprocal space), a further difference is the choice of the free variable, which is the constraining field ${\bf B}^{\text{C}}_{i}$ in our method, while in Grotheer's et al. approach~\cite{Grotheer:2001db,Grotheer:2000gp} it is the cone-angle. The transverse field method has a better scaling with respect to the size of the system than the frozen magnon method has, since the number of determined variables is equal to the number of magnetic atoms for each calculation, while in the latter method only a single variable is determined in each calculation. While the constraining fields in the transverse field method are converged self-consistently, which is computationally more demanding compared to non-self consistent methods, this is compensated well by the fact that the constraining fields are robust quantities that are usually much faster to converge with respect to the computational parameters than small total energy differences.~\cite{Kurz} 

\subsection{Beyond the Heisenberg model}

The regular Heisenberg model performs poorly when large deviations from the ground state are considered for itinerant magnetic systems in the sense that the convergence of the residuals with respect to the number of parameters is slow compared to more complex models.~\cite{Singer}. In the final section, the good scaling of the constraining field method is therefore applied to parameterise models -- that besides the bilinear terms -- also contain higher order interactions such as bi-quadratic, three- and four-spin interactions. The exchange fields of two models are considered here in addition to the regular Heisenberg exchange field  ${\bf H}_{0}$ (Eq.~\ref{eq:H0}). First, ${\bf H}_{0}$ together with bi-quadratic, three- and four-spin interactions, called ${\bf H}_{2}$:%
\begin{equation}
\hspace{-0.1cm}{\bf H}_{2i}=-\frac{1}{m_{i}}\left(\sum_{j}J_{ij}{\bf e}_{j} +\frac{1}{2} \sum_{j,k \neq l}B_{ijkl}({\bf e}_{j})({\bf e}_{k}\cdot {\bf e}_{l})\right)
\label{eq:H2}
\end{equation}%
and then ${\bf H}_{2}$ restricted to bi-quadratic interactions only, called ${\bf H}_{1}$:

\begin{equation}
{\bf H}_{1i}=-\frac{1}{m_{i}}\left(\sum_{j}J_{ij}{\bf e}_{j} +\sum_{j}B_{ijji}({\bf e}_{j}) ( {\bf e}_{i} \cdot {\bf e}_{j})\right)
\label{eq:H1}
\end{equation}%
 where $J_{ij}$ and $B_{ijkl}$ are the exchange parameters. Similarly to Eq.~\eqref{eq:susc2}, the parameters can be related to the constraining fields. For example, applied to the most general model ${\bf H}_{2}$ (Eq.~\ref{eq:H2}), this gives:  

\begin{equation} 
\begin{split}
{\bf B}^{\text{C}}_{i}\cdot {\bf B}^{\text{C}}_{i} & =  \frac{1}{m_{i}}\sum_{j}J_{ij}{\bf B}^{\text{C}}_{i} \cdot {\bf e}_{j}\\
& + \frac{1}{2m_{i}} \sum_{j,k\neq l}B_{ijkl}({\bf B}^{\text{C}}_{i} \cdot {\bf e}_{j} )({\bf e}_{k}\cdot {\bf e}_{l}).
\end{split}
\label{highereq}
\end{equation}

In order to address the problem of overfitting and access the predictive power of the models, a leave-one-out cross-validation analysis is performed. Here, each single data point is left out in turn from the set, and the parameters are then extracted from the remaining set of data points and used to predict the value of the left-out data point. The predicted mean squared error (P-MSE) obtained from the differences between the predicted and actual data points provides a better measure of the predictive power of the models than the regular MSE. While the MSE decreases more or less monotonously with respect to the number of parameters regardless of model and material under study, the P-MSE has a global minimum that marks the onset of overfitting. 

In the regular Heisenberg model, the terms are usually ordered and included in the model according to the relative distances of the atomic pairs. For more complicated models, additional conventions have to be made. In ${\bf H}_{1}$ and ${\bf H}_{2}$ the higher order terms are ordered according to the sum of all the relative distances between the members of the moment-quadruples. Hence, the nearest neighbour bi-quadratic interaction is always the first higher order term included in the model. In addition, rules are introduced that order the $J_{ij}$ with respect to the $B_{ijkl}$ coefficients. For ${\bf H}_{1}$ let $J_{ij} > B_{ijji}$  if the sum of all the relative distances between the members of the moment-quadruple that define the $B_{ijji}$ is less than four times the distances between the pair that defines the $J_{ij}$. For ${\bf H}_{2}$ the factor four is reduced to a factor two. Essentially the two dimensional problem with possible different length cutoffs for bilinear and higher order interactions is reduced to one dimension by the use of  a fixed ratio. The drawback of this simplification is of course that the possible predictive power of the more complex models might be underestimated. These factors may be optimised and made specific to the material and the chosen sampling of non-collinear states using the cross validation procedure. However an optimisation using our data on bcc-Fe resulted in similar ratios between the cutoff lengths (four and two), so these rules are used in this work in order to simplify the comparisons to the standard Heisenberg model. In general it is recommended to do the full model optimisation since it may identify the cases where the ground state is stabilised by higher order interactions and will further increase the predictive power of the model.


\subsection{Computational Details}

The main results presented in this paper are obtained with the new formalism described above as implemented in the full potential APW+lo Elk code~\cite{httpelksourcefor:0ZCfU9-K}. The computational details for the three systems bcc Fe, fcc Ni and FeCo are the same, except for the number of {\bf k}-points. The local spin density approximation (LSDA) functional by Perdew and Wang~\cite{Perdew-Wang} is used throughout this work. For the band energy calculations a $31\times  31\times31$ {\bf k}-point mesh is used for fcc Ni and FeCo while a  $41\times 41\times 41$ {\bf k}-point mesh is used for bcc Fe. The constraining fields are obtained using a  $15\times 15\times 15$ {\bf k}-point mesh for fcc Ni and FeCo, while a  $21\times 21\times 21$ {\bf k}-point mesh is used for bcc Fe. The muffin-tin radius was set to 1.23 \AA ~for all atoms. The angular cutoff for the APW functions, for the muffin-tin potential and the density was set to 9. The maximum length of the \{{\bf G} + {\bf k}\}-vectors was regulated by fixing its product with the average muffin-tin radius to 9. The maximum length of the \{{\bf G}\}-vectors describing the interstitial potential and the density was set to 9.5 \AA$^{-1}$.  We utilise unrestricted intra-atomic non-collinearity in the calculations. The experimental lattice constants of 2.87 \AA,  3.50 \AA ~and 2.83 \AA ~were considered for bcc Fe, fcc Ni and FeCo, respectively. 
%

For the calculations of exchange parameters 100 spin-spirals were considered with randomly generated wave-vectors for bcc Fe, while 90 spin-spirals were considered for fcc Ni and FeCo. Cone-angles of 0.25 rad on one or two magnetic sub-lattices were used for all the results in this paper, while angles up to 0.5 rad were considered to ensure that our results did not depend sensitively on the chosen angles. The constraining fields were considered converged when the change of the constraining fields between two successive iterations was less than 0.5~\%. For FeCo we also calculate the exchange parameters using the Korringa-Kohn-Rostocker (KKR) method and the MFT method as developed by Lichtenstein et al.~\cite{Liechtenstein:1987br} and implemented in the M\"{u}nich SPR-KKR band structure program package~\cite{Munich}. The disordered local moment (DLM) model  has been employed for describing the paramagnetic state of Fe~\cite{gyorffy1985, pindor1983}. This model treats spin-up and spin-down components in equal concentration, assuming a completely random distribution of magnetic moments for each magnetic element, in the sense that all correlations are absent. The electronic structure of the DLM-state has been obtained using the coherent potential approximation (CPA)~\cite{soven1968,taylor1967}, which accurately describes disordered systems in the single-site approximation.

In order to sample a large variety of non-collinear states for the different beyond-Heisenberg models and to avoid issues of linear dependency between higher order parameters, the cubic 16-atom unit cell was considered for bcc Fe. Here 125 {\bf k}-points were evaluated and otherwise the same set of computational parameters used previously, was employed. The constraining fields and magnetic moments were converged for five sets of magnetic configurations. The cone angles $\theta_{i}$ for the 16 atoms were randomised in the intervals $(n-1)\pi/10 < \theta_{i} < n\pi/10$ for the five sets defined by $1 \le n \le 5$. Furthermore, the phase factors $\phi_{i}$ were randomised between $0$ and $2\pi$ and a spin-spiral with a random wave-vector applied for each configuration. With increasing $n$  increasingly disordered states are considered from the almost collinear states of $n=1$ to the almost paramagnetic $n=5$. In total, 100 different magnetic configurations were considered. 

\section{Results}

In order to evaluate the accuracy of the spin-spiral total energy differences using the MFT with and without the use of constraining fields, spin-spiral total energies calculated fully self-consistently were used as a benchmark. 
Three different situations were considered for the calculation of total energies: (i) non-self-consistent calculations without any constraining fields (referred in the following as {\it NSC} and marked with blue circles); (ii) non-self-consistent calculation with constraining fields (referred to as {\it NSC-F}, green triangles) and (iii) self-consistent calculations (referred to as {\it SC}, red squares).  Only for bcc Fe and fcc Ni are the spin-spiral dispersions directly related to the adiabatic magnon dispersion through a scaling factor since they -- unlike FeCo -- have a single magnetic sub-lattice. 

\subsection{Bcc Fe}
\label{bccFe}
\begin{figure}[htb]
\includegraphics[width=0.441\textwidth]{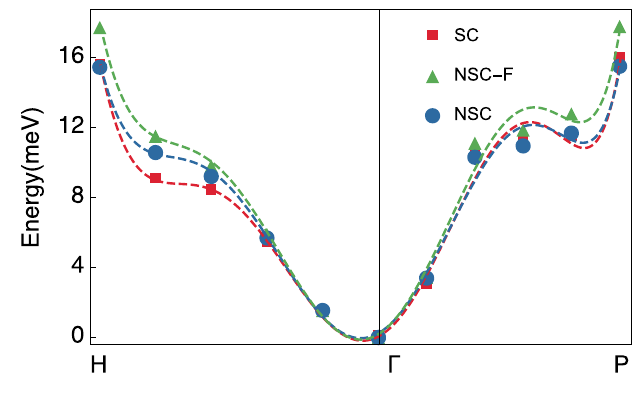}
 \caption{(color online) Spin-spiral dispersion in bcc Fe. The blue circles and green triangles represent non-self-consistent total energies calculated without- (blue) or with (green) constraining fields. The red squares represent self-consistent total energies. }
\label{fig1}
\end{figure}

We calculated the spin-spiral energy dispersion in bcc Fe along the H--$\Gamma$--P direction (Fig.~\ref{fig1}) for all the three cases mentioned before: \textit{NSC}, \textit{NSC-F} and \textit{SC}.
The magnetic field ${\bf B}_{eff}({\bf r})$ is aligned closer to the z-axis when the constraining fields are neglected and thus, it corresponds to a situation of smaller cone-angles. This results in lower total energy differences with respect to ${\bf q}=0$ and it is why the {\it NSC-F} energies lie above the {\it NSC} case. 

The behaviour is consistent with the results of previous studies where quantities such as exchange parameters, magnon energies and the Curie temperatures $T_{C}$ are increasing when the MFT calculations are corrected for the LWA.~\cite{antro-shilf,Bruno:2003ir,Antropov,Katsnelson:2004ha} Indeed, we also obtained a higher $T_{C}$ for bcc Fe with the corrected frozen magnon method than with the conventional one as can be seen in Table~\ref{TC}.

\begin{table}[htb]
\caption{The Curie temperatures ($T_{C}$) are listed in K and the theoretical results are all obtained in the mean-field approximation that typically overestimates the $T_{C}$ by $\sim$ 15 \%. The exchange parameters of Ref.~\cite{shilf-antrop} are obtained using the MFT approach of Liechtenstein et al.~\cite{Liechtenstein:1987br} in the LWA.}
\vspace{0.25cm}

\begin{ruledtabular}
\begin{tabular}{c c c c c c}
 & {\it NSC}  & {\it NSC-F} &  {\it TF} & Ref.~\cite{shilf-antrop} & Exp.\\ [0.5ex] 
\hline
Bcc Fe &1067&1143&1077 & 1050 & 1043\\
Fcc Ni&351&572&533 & 406 & 627 \\
FeCo &1853& 2261 &1934 & - & -\\
\end{tabular}
\end{ruledtabular}
\label{TC}
\end{table}

It is worth noting that we have intra-atomic collinearity in our non-self consistent calculations for bcc Fe (and for fcc Ni) since we have only a single magnetic sub-lattice in this case and therefore, have a collinear $\Gamma$-point reference state. In a previous study,~\cite{antro-shilf} it was shown that suppressing intra-atomic non-collinearity in bcc Fe gives rise to an increase of the calculated $T_{C}$ when derived from self-consistent spin spiral total energies. This observation is consistent with our results since it suggests that the non-self consistent calculations -- where we have intra-atomic collinearity -- should get higher total energy differences compared to the self consistent calculations. This may explain why the agreement with self-consistent spin-spiral total energies decreases somewhat with the introduction of the fields in the non-self-consistent calculations. This is likely due to the fact that the error cancellation is lost between the neglect of the fields (that lowers the effective cone-angles and thus energies) and the imposition of intra-atomic collinearity (that increases the energies of the spin-spirals). 

\begin{figure}[htb]
\begin{center}
\includegraphics[width=0.441\textwidth]{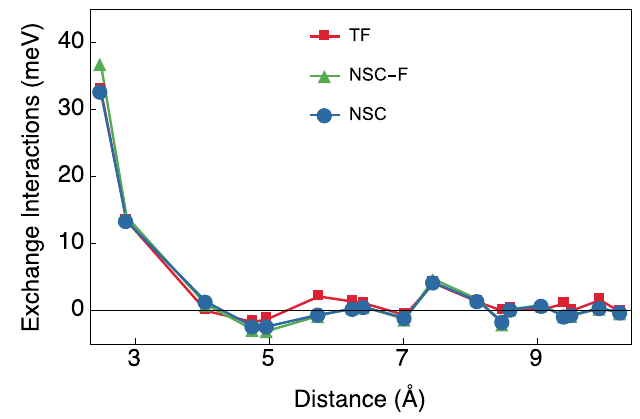}
\caption{(color online) Exchange parameters for bcc Fe. The blue circles and green triangles represent exchange parameters calculated non-self-consistently without (blue) and with (green) constraining fields. The red squares represent exchange parameters calculated with the transverse field method.}
\label{fig2}
\end{center}
\end{figure}
The most pronounced difference in the exchange parameters is the larger nearest neighbour interaction of the {\it NSC-F} case compared to the {\it NSC} case (Fig.~\ref{fig2}). This difference is of similar size to the one obtained in previous studies between methods that worked in the LWA and those that went beyond, using the linear response theory~\cite{Antropov}  or self-consistent spin-spirals.~\cite{antro-shilf} Similarly, the error cancellation can explain the excellent match obtained between self-consistently converged exchange parameters and the exchange parameters from previous studies using the MFT~\cite{Bergqvist:pdLfo1Fw,Frota-Pessoa,Kvashnin:2015uj}. 
In fact, an almost perfect match is obtained with the results of Bergqvist~\cite{Bergqvist:pdLfo1Fw}. It should be noted, however, that bcc Fe is not very well described in a strict Heisenberg model. The exchange parameters are configuration-dependent and specifically the ratio between the dominating nearest- and second nearest neighbour interactions is known to depend sensitively on the cone-angle of the spin-spirals.~\cite{kubler} Some differences between the results of various calculations available in the literature can be expected for that reason. To avoid the configuration-dependence, a more complex Hamiltonian has to be considered.~\cite{Singer}  This issue will be discussed in further detail later in the article. 

\subsection{Fcc Ni}
\label{fccni}

\begin{figure}[htb]
 
\includegraphics[width=0.441\textwidth]{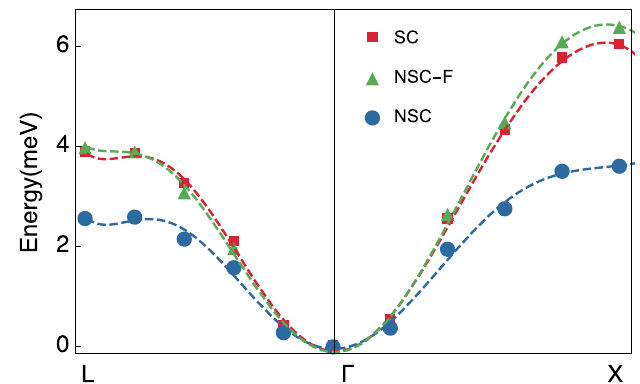}
 \caption{(color online) Spin-spiral dispersion in fcc Ni.  The blue circles and green triangles represent non-self-consistent total energies calculated without- (blue) or with (green) constraining fields. The red squares represent self-consistent total energies. Dashed lines represent fitted polynomials.}
\label{fig3}
\end{figure}

Introducing the pre-converged constraining fields into the non-self consistent calculations gives results that match excellently with self-consistent results while the conventional frozen magnon method produces a dispersion curve that is significantly lower (Fig.~\ref{fig3}).  

\begin{figure}[htb]
\includegraphics[width=0.441\textwidth]{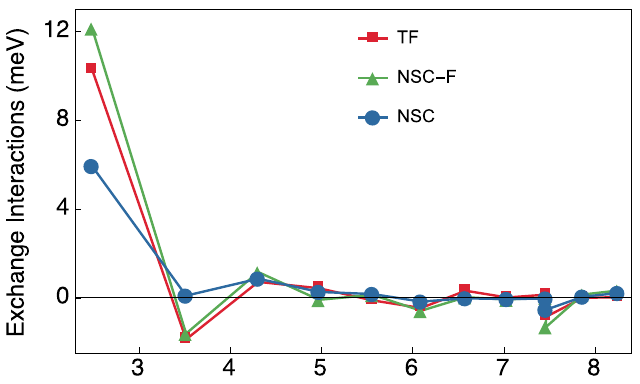}
\caption{(color online) Exchange parameters in fcc Ni. The green triangles and blue circles represent exchange parameters calculated non-self-consistently with- (green) or without (blue) constraining fields. The red squares represent exchange parameters calculated with the transverse field method.}
\label{fig4}
\end{figure}

This means that correcting for the LWA removes almost entirely the large differences between non-self consistent and self-consistent results.  As a consequence, the exchange parameters of the corrected magnon method correspond more closely to the exchange parameters obtained by the transverse field method, shown in Fig.~\ref{fig4}. We note that there is a substantial difference between the size of the nearest neighbour interaction obtained in our study when derived from the transverse field or the corrected frozen magnon method and previous self-consistent results.~\cite{antro-shilf}. We also get substantially weaker nearest neighbour interaction compared to results of linear response theory.~\cite{Antropov} It is however clear from Fig.~\ref{fig3} that the {\it NSC-F} energy differences are close to the self-consistent results, which indicates that the mapping of the self-consistent total energy landscape is correct, at least for the high symmetry lines probed here. This is further reinforced by the close agreement between the exchange parameters calculated self-consistently from the transverse field and the ones obtained from the corrected frozen magnon method. We also note that our results for the $T_{C}$ give a decent match to previous time-dependent density functional simulations~\cite{A-SpinD}.

The $T_{C}$ of fcc Ni is well known to be underestimated by MFT calculations that operate in the LWA and employ a LSDA functional.~\cite{Antropov,Bruno:2003ir, Halilov, antro-shilf, shilf-antrop} The transverse field method and corrected frozen magnon method give a much improved prediction of the $T_{C}$, as can be seen in Table~\ref{TC}, but still fall significantly short of the experimental $T_{C}$. Constraining fields were also calculated for configurations with larger angles and it was found that the dominating nearest neighbour interaction tends to weaken with the onset of magnetic disorder, while no other interactions grew substantially in size. This indicates that a more thorough sampling of non-collinear states will rather decrease than increase the predicted transition temperature. The discrepancy between theory and experiments should be sought elsewhere. This further strengthens the case that the LSDA functionals underestimate the exchange interactions or that non-adiabatic processes and longitudinal fluctuations not captured by the Heisenberg model also play an important role in determining the  $T_{C}$ in fcc Ni.~\cite{A-SpinD, longi}.

\subsection{ B2 Structured FeCo}
\label{feco}

\begin{figure}[t!]
\center{(a) Fe and Co moments tilted }
 \subfloat 
   {\label{fig:fig5a}%
\vspace{3cm}
     \includegraphics[width=0.95\linewidth]{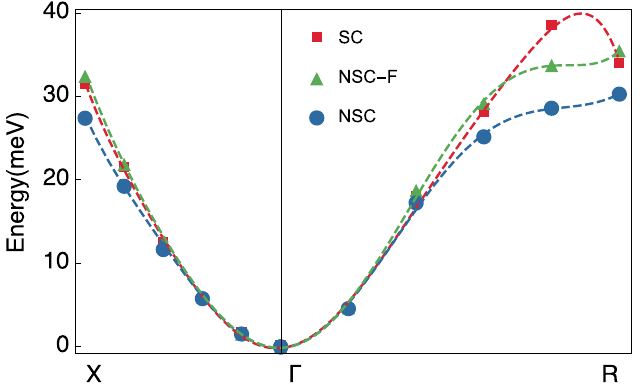} }
\vspace{-0.5cm}
\newline
\center{(b) Fe moments tilted} 
  \subfloat 
   {\label{fig:fig5b}%
      \includegraphics[width=0.95\linewidth]{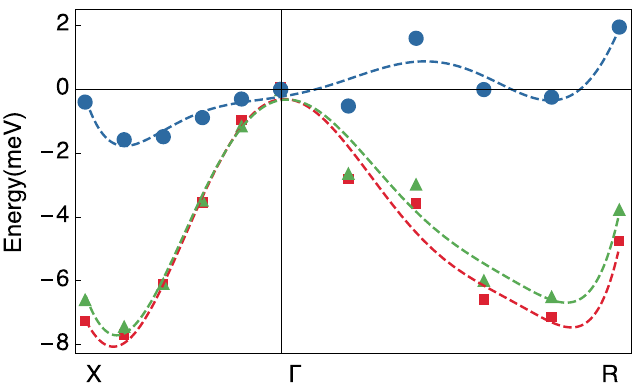} }
\vspace{-0.5cm}
\newline
\center{(c) Co moments tilted} 
 \subfloat 
  {\label{fig:fig5c}%
   \includegraphics[width=0.95\linewidth]{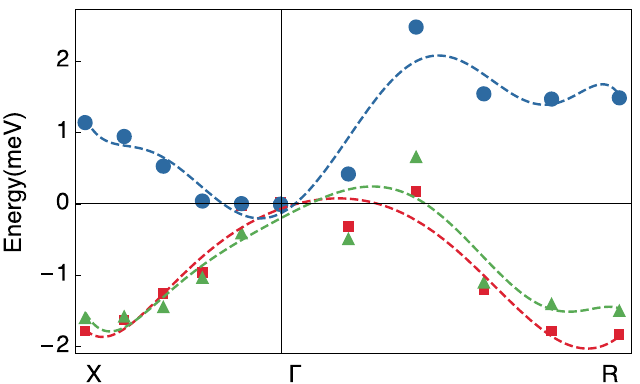}   } 
\vspace{-0.250cm}
\caption{(color online) Spin-spiral dispersion curves in FeCo, when: (a) when both the magnetic moments on the Co and Fe sub-lattices are tilted, (b) with only the magnetic moments on the Fe magnetic sub-lattice tilted or (c) with only the magnetic moments on the Co magnetic sub-lattice are tilted. The blue circles and green triangles represent non-self-consistent total energies calculated without- (blue) or with (green) constraining fields. The red squares represent self-consistent total energies. Negative values occur here since the gamma point is a non-collinear state. Dashed lines represent fitted polynomials.}
 \label{fig:fig5}
\end{figure}

For FeCo two different cases of magnetic configurations were considered for the dispersion: (i) moments are only titled on one magnetic sub lattice (either Fe or Co and presented in figs.~\ref{fig:fig5c} and~\ref{fig:fig5b}), while the orientations of the moments on the other sub-lattice are kept fixed and (ii) moments are tilted simultaneously on both magnetic sub-lattices (Fe and Co and presented in Fig.~\ref{fig:fig5a}).
\\ 

\begin{figure}[htb!]
\center{(a) Fe-Co Exchange}
 \subfloat 
  {\label{fig:fig6a}%
   \includegraphics[width=0.95\linewidth]{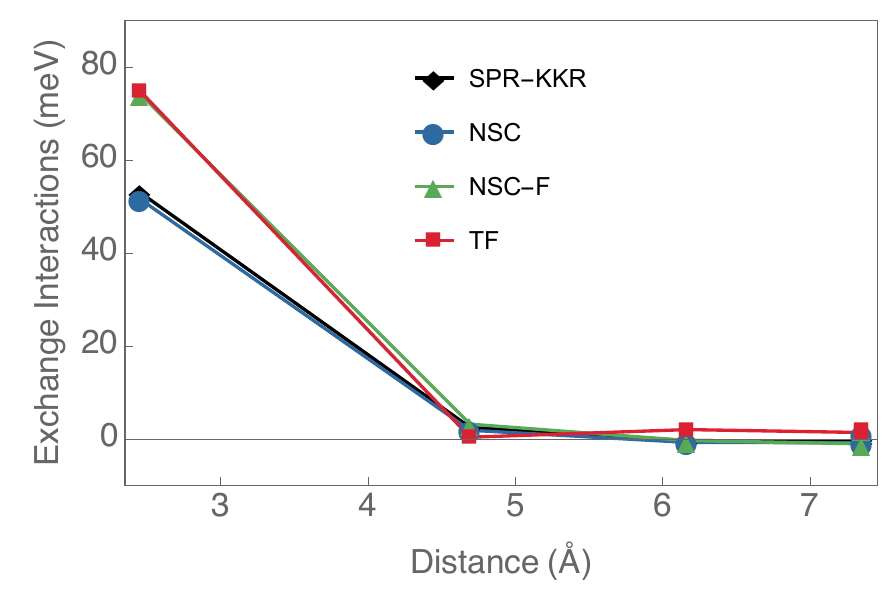}  }
\newline
\vspace{-0.45cm}
\center{(b) Fe-Fe Exchange}
 \subfloat 
  {\label{fig:fig6b}%
   \includegraphics[width=0.95\linewidth]{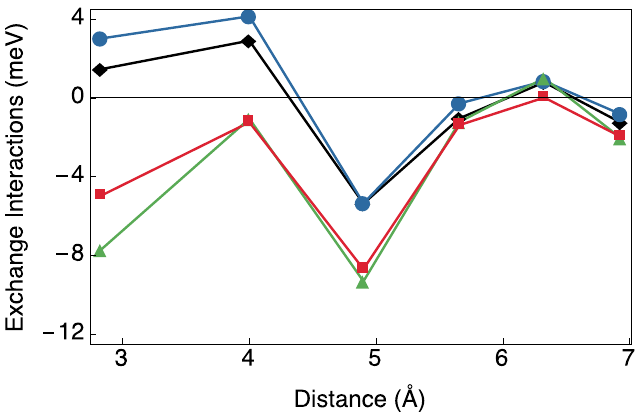}  }  
\vspace{-0.45cm}
\newline
\center{(c) Co-Co Exchange}  
 \subfloat 
  {\label{fig:fig6c}%
   \includegraphics[width=0.95\linewidth]{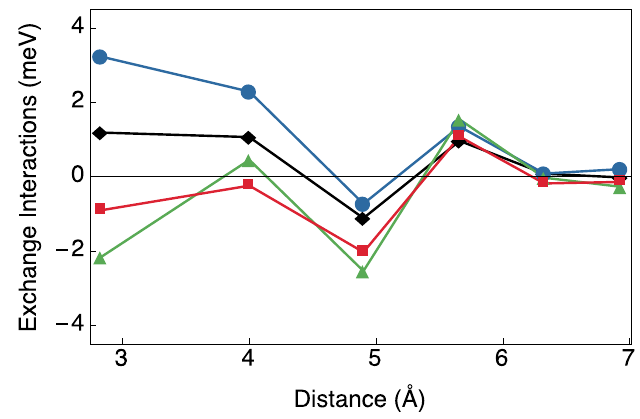} }      
\caption{(color online) Exchange interactions between: (a) the two magnetic sub-lattices, Fe and Co, (b) within the Fe magnetic sub-lattice and (c) within the Co sub-lattice. The green triangles and blue circles represent exchange parameters calculated non-self-consistently with (green) and without (blue) constraining fields. The red squares and black diamonds represent exchange parameters calculated with the transverse field method and the method of Lichtenstein et al.~\cite{Liechtenstein:1987br}, respectively. 
} 
\label{fig:fig6}
\end{figure}


The {\it NSC-F} dispersion curves correspond considerably more accurately to the {\it SC} case than the {\it NSC} for FeCo. The difference becomes especially pronounced qualitatively when only the Co (Fig.~\ref{fig:fig5c}) or the Fe (Fig.~\ref{fig:fig5b}) sub-lattice is tilted even though it needs to be pointed out that the energy scale is much smaller in this case compared to the case where both magnetic sub-lattices are tilted. The energy differences of the spin spirals in figs.~\ref{fig:fig5b} and~\ref{fig:fig5c} do not depend on the inter-lattice exchange parameters since the angles between atomic moments belonging to different sub-lattices are constant. Hence we can expect significant differences in the intra-lattice exchange parameters. Indeed the {\it NSC-F} intra-lattice parameters, shown in Figs.~\ref{fig:fig6b} and \ref{fig:fig6c}, correspond much closer to the exchange parameters obtained by the transverse field method than the {\it NSC}-exchange parameters obtained by the conventional MFT calculations. Also the nearest neighbour Fe-Co interaction is significantly stronger when obtained by the methods that goes beyond the LWA. It can be noted that the energy scales in Figs.~\ref{fig:fig6a}, \ref{fig:fig6b} and \ref{fig:fig6c} here are similar to the corresponding graphs in Figs.~\ref{fig:fig5a} ,~\ref{fig:fig5b} and ~\ref{fig:fig5c}. Exchange parameters were also obtained using a KKR method employing the approach of Lichtenstein et al.~\cite{Liechtenstein:1987br}  and found to be similar to the {\it NSC} calculations (the purple vs the green symbols in figs.~\ref{fig:fig6a}, \ref{fig:fig6c} and \ref{fig:fig6b}). Also previous KKR calculations by MacLaren et al.~\cite{Anonymous:1999fz} match our KKR results well. This shows that the conventional MFT calculations fail to describe the self-consistent energy landscape of FeCo due to the LWA regardless of the specifics of the implementation. In this case there is no experimental value to compare to since FeCo undergoes a structural phase transition before losing the magnetic ordering. 

\subsection{Beyond the Heisenberg model}

\label{mse}

To assess the predictive power a cross validation analysis is employed and it is shown that the extended models give a significantly increased accuracy for bcc Fe.
%
%
\begin{figure}[htb]
\subfloat
 {\label{fig:fig7a}%
  \includegraphics[width=0.95\linewidth]{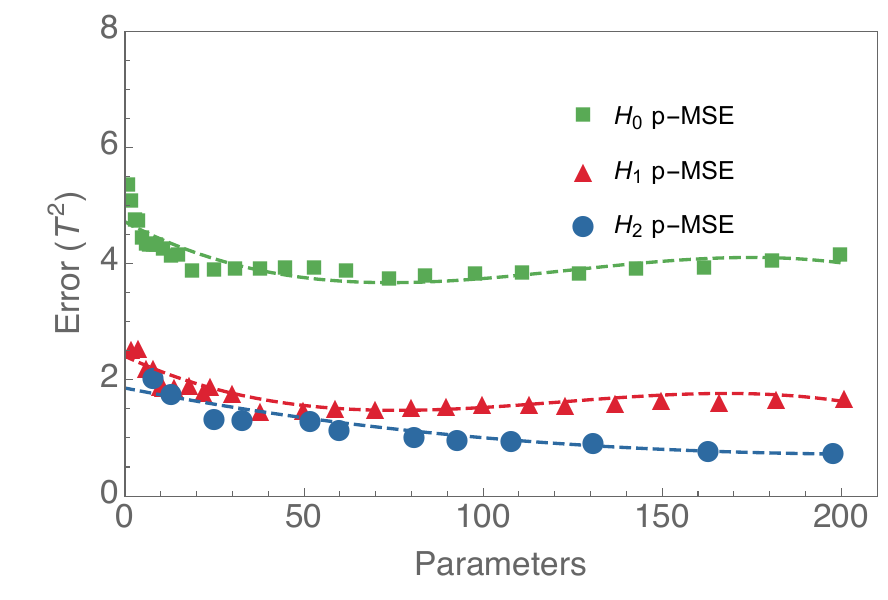} }
\vspace{-0.40cm}
\newline 
\subfloat
 {\label{fig:fig7b}%
  \includegraphics[width=0.95\linewidth]{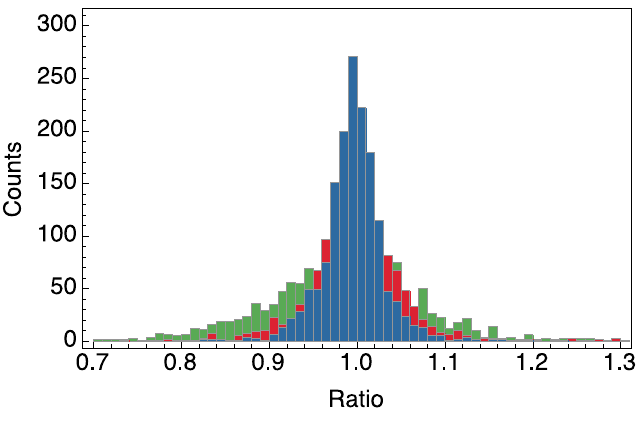} }
\vspace{-0.35cm}
\caption{(color online) (a) The mean squared differences between ${\bf B}^{\text{C}}_{i} \cdot {\bf B}^{\text{C}}_{i}$ as obtained from direct {\it ab initio} calculations and  ${\bf B'}^{\text{C}}_{i} \cdot {\bf B'}^{\text{C}}_{i}$ is obtained from the parameterised models derived from the data-set excluding ${\bf B}^{\text{C}}_{i}$. The regular Heisenberg model is labeled $H_{0}$, $H_{0}$ with bi-quadratic interactions added is labeled $H_{1}$ and $H_{1}$ extended by three- and four-spin interactions is labeled $H_{2}$. (b) The ratio ${\bf B}^{\text{C}}_{i}/{\bf B'}^{\text{C}}_{i}$, where ${\bf B'}^{\text{C}}_{i}$ is obtained from the parameterised models derived from the data-set excluding $ {\bf B}^{\text{C}}_{i}$. The color scheme in (b) is the same as in (a) for the different models.}
\label{fig:fig7}
\end{figure}
As seen in Fig.~\ref{fig:fig7a} it is clear that the extended Heisenberg models are significantly superior to the regular Heisenberg model and that the accuracy of the models follow the expected complexity order, with ${\bf H}_{2}$ more accurate than ${\bf H}_{1}$ and ${\bf H}_{1}$ more accurate than ${\bf H}_{0}$. In Fig.~\ref{fig:fig7a} the global minimum of the P-MSE of the ${\bf H}_{2}$-model is not clear contrary to the cases of ${\bf H}_{0}$ and ${\bf H}_{1}$. However if the number of parameters is increased beyond 200 it is found that 198 parameters is the global minimum using the fixed ratios between the cutoff lengths for the bilinear and higher order terms. The result of Singer et al. ~\cite{Singer} that a single bilinear and bi-quadratic parameter give a better fit than any number of bilinear parameters is reproduced as seen by comparison between ${\bf H}_{1}$ and  ${\bf H}_{0}$, although their analysis is based on MSE rather than the P-MSE and a different sampling of non-collinear states. It is interesting to note that the predictive power of the multi-spin model ${\bf H}_{2}$ is significantly improved compared to ${\bf H}_{1}$ by the inclusion of hundreds of small three- and four-spin interactions. When the ratios are optimised it is found that the P-MSE cannot be significantly improved within cutoff-limits that give a parameterisation from the available data. It is possible that the optimisation for a more extensive collection of data would make a larger difference between optimised and fixed ratios. 
%
The ratio between the predicted and directly calculated size of the of the transverse exchange fields are shown in Fig.~\ref{fig:fig7b} to make it easier to evaluate the accuracy the different models. Here small fields might of course have a ratio significantly different from one and still not contribute with a large absolute error. But an overall sense of the accuracy for the bulk of the data points can nevertheless be obtained. For the extended models the fields are predicted within  $10\%$ of the directly calculated fields and for the regular Heisenberg model the fields are predicted within $20\%$. The ratios have been evaluated with a number of parameters determined by the minimum of the P-MSE, i.e. 74, 38 and 198 parameters for ${\bf H}_{0}$, ${\bf H}_{1}$ and ${\bf H}_{2}$ respectively. The variations in the sizes of the magnetic moments were less than $15\%$ for the data-set and insignificant for the smaller angles. It was investigated wether the Hamiltonians could be improved by including a bilinear scaling with respect to the amplitudes of the moments. A version of every Hamiltonian where the parameters where multiplied by the moments ${\bf m}_{j}$ rather than the directional vectors ${\bf e}_{j}$ were tested and found to give significantly worse accuracy for all cases, perhaps contrary to expectations. A more thorough investigation of this matter is postponed to the future.   

 It is found that the DLM results match the self-consistent results for bcc Fe perfectly when employing the regular Heisenberg model. The regular Heisenberg model doesn't capture features in the local spin density approximation (LSDA) total energy landscape that depends upon this particular use of the coherent potential approximation (CPA). Those features may be described by the higher order interactions and most importantly the strong bi-quadratic nearest neighbour interaction.
  
In the work by Szilva et al.~\cite{szilva2013}, they derived collinear exchange parameters for the bilinear and bi-quadratic spin Hamiltonian. Their obtained values in the case of bcc Fe compare very well qualitatively to our calculated exchange interactions for the same system, when using ${\bf H}_1$ model. It is expected that the two approaches give some quantitative differences since our approach is self-consistent while theirs is non-self consistent and since we also sample states far away from the collinear state when parametrising the model.

\begin{figure}[hbt]
 \subfloat {  
   \includegraphics[width=0.49\textwidth]{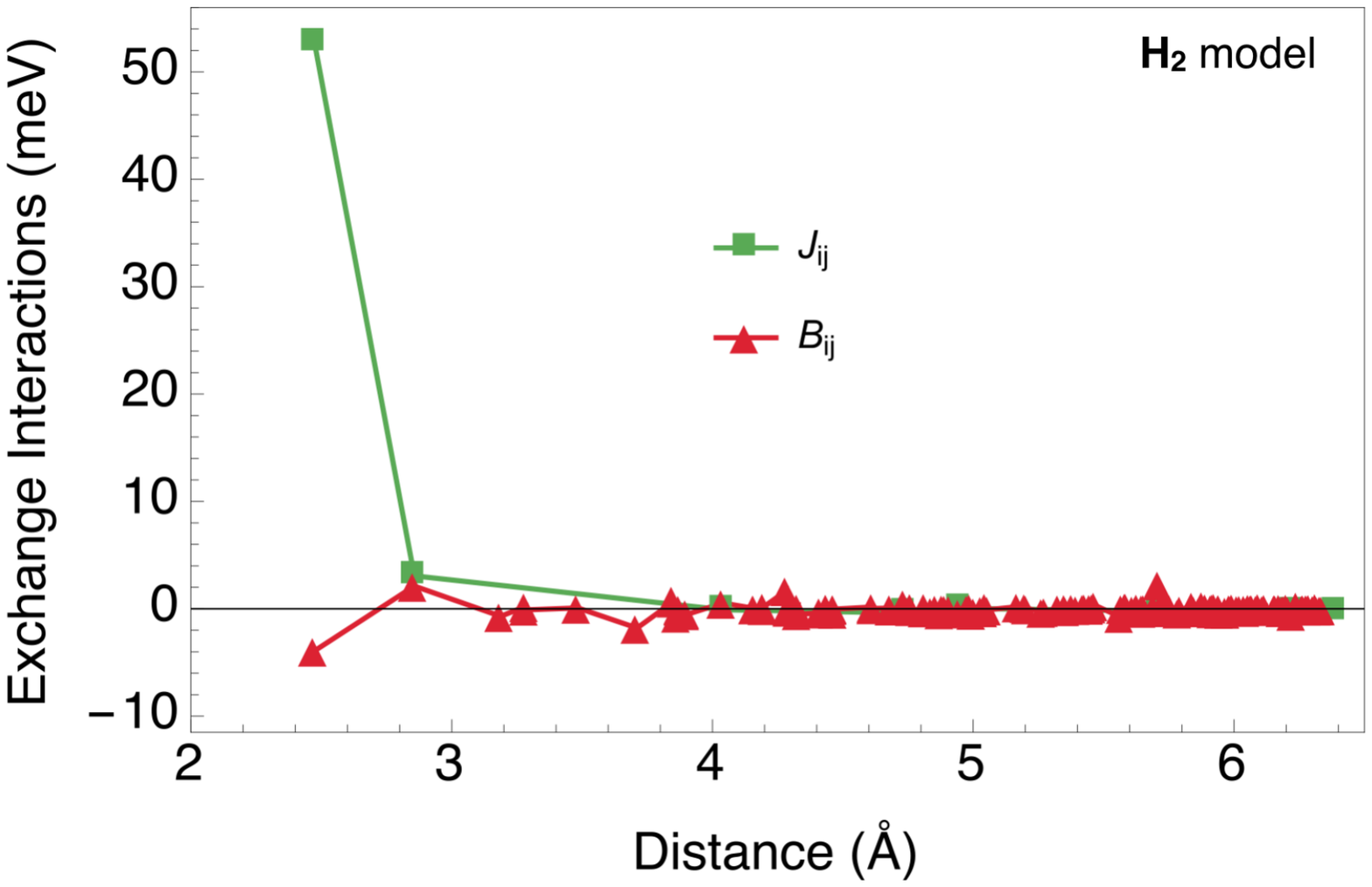}  
\label{H2}}
\newline
\vspace{-1.2cm}
  \subfloat{
 \includegraphics[width=0.49\textwidth]{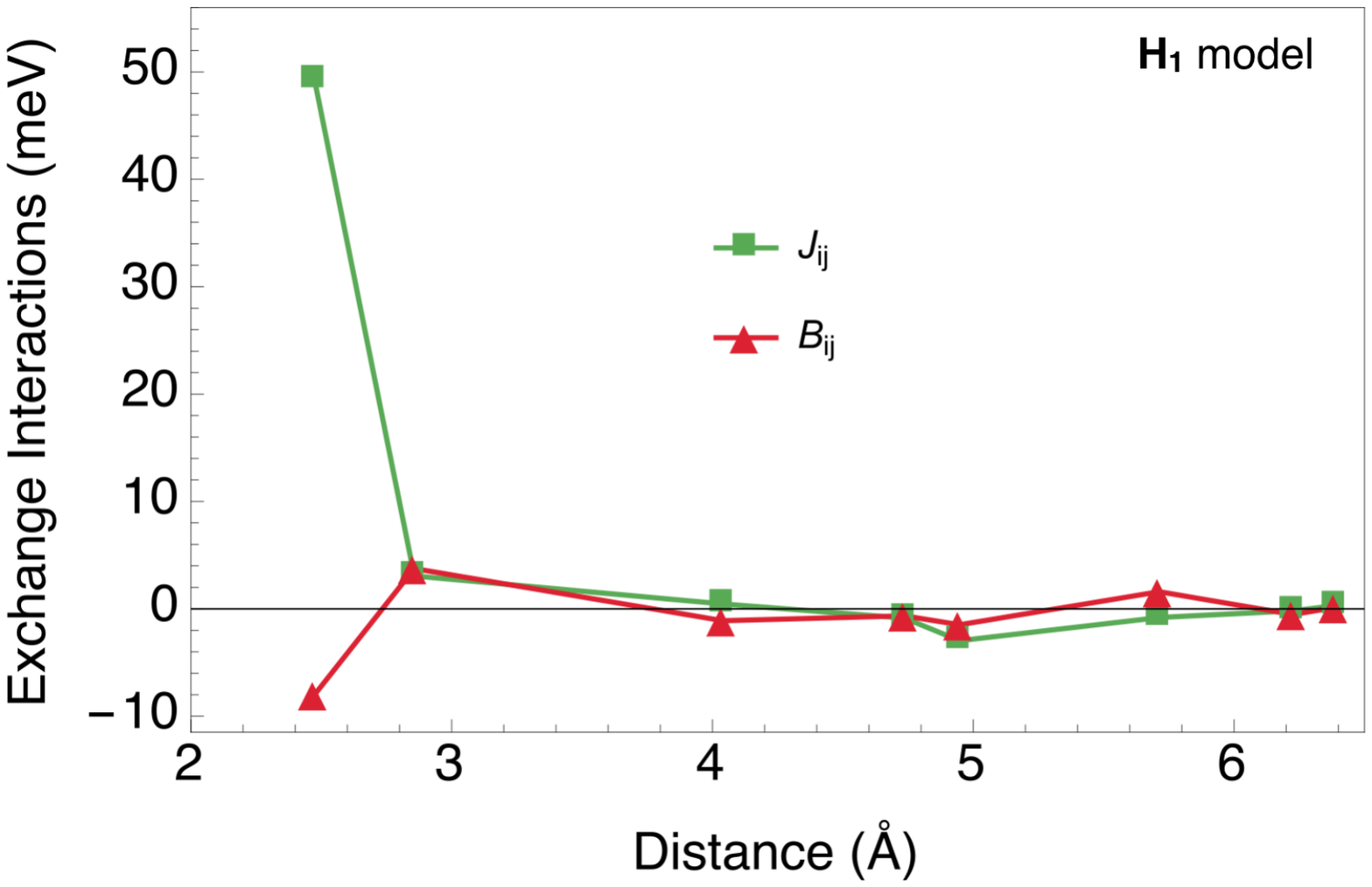}    
\label{H1}}
\vspace{-0.8cm}
\newline
    \subfloat{
    \includegraphics[width=0.49\textwidth]{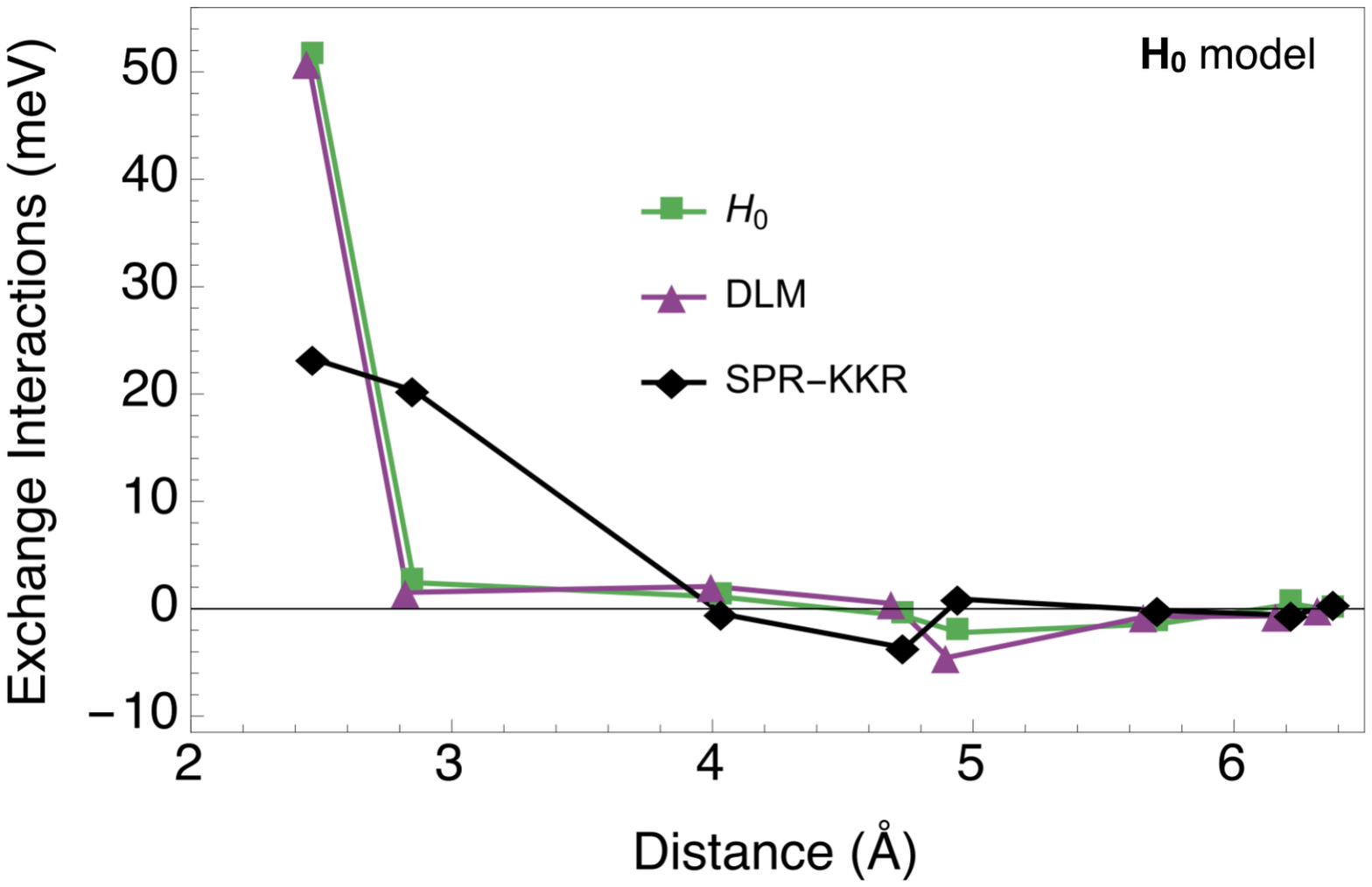} 
\label{H0}}
\vspace{-0.5cm}
\caption{(color online) The parameterisations for the three models ${\bf H}_{2}$, ${\bf H}_{1}$ and ${\bf H}_{0}$ are shown. Here the sums of the distances between the moment-quadrupoles are divided by four for ${\bf H}_{2}$ and ${\bf H}_{1}$ in order to make the distances of the exchange parameters $J_{ij}$ comparable with the $B_{ijkl}$. For the ${\bf H}_0$ model, the exchange parameters are calculated using Lichtenstein et al.~\cite{Liechtenstein:1987br} expression as implemented in a KKR-code, with and without the DLM formalism.} 
\label{fig:Models}
\end{figure}

\section{Conclusions}

Much of the discussions regarding the accuracy of the exchange parameters determined with the MFT has revolved around the results for bcc Fe. This is a material where conventional MFT calculations in the LWA and self-consistent total energy calculations agree very well for small deviations in the magnetic structure and where the introduction of any correction scheme therefore has little room for improving the results. This excellent agreement can be understood to be partly due to the error cancelation between (i) the neglect of the constraining fields and (ii) the assumption of intra-atomic collinearity in the calculations. Therefore, the bcc Fe exchange parameters of the transverse field method agree very well with the results of calculations using the MFT in the LWA, while the corrected frozen magnon method gives slightly different quantitative predictions for the exchange parameters. But for any material and state where ${\bf B}^{C}({\bf r}) \gtrsim {\bf B}^{XC}({\bf r})$, significant differences can be expected between non-self-consistent and self-consistent results. This is the case for the spin-spiral total energy calculations for fcc Ni and  FeCo, where the conventional use of the MFT doesn't result in a correct description of the total energy as a function of the directions of the moments. The neglect of the contributions of the constraining fields to the potential is crucial for this discrepancy as we can see from the comparison between our two different frozen magnon methods. 

We can see that going beyond the LWA through the use of the corrected frozen magnon- or transverse field method does result in a critical temperature that is significantly closer to experiments for fcc Ni using LSDA. When larger angles were applied, the exchange tended to weaken which indicates that either the LSDA itself fails to give an accurate description of fcc Ni or a more complex treatment beyond the Heisenberg model is needed. This finding is consistent with previous studies that go beyond the LWA. Besides, turning to more accurate electronic structure methods, it would be interesting to see whether the theoretical predictions could be made more accurate by introducing longitudinal fluctuations of the magnetic moments into the effective Hamiltonian.  

The corrected frozen magnon method gives an overall accurate description of the self-consistent energy landscape of all the three materials covered in this study, while the conventional method only works well for bcc Fe. This indicates that non-self consistent spin-spiral calculations are, in general, not the most efficient approach for the purpose of parameterising the total energy of non-collinear states in itinerant magnetic systems. This is due to the fact that the constraining fields that remove the differences to self-consistent total energies need to contain information of the parameterisation itself. However, the procedure can still make sense if the constraining fields are not fully self-consistently converged. In our corrected frozen magnon method the potential is kept fixed while the constraining fields are converged. This results in computational costs for the corrected frozen magnon method comparable with the transverse field method for the small magnetic systems considered in this study, while the transverse field method scales better and therefore is clearly favourable to use for larger magnetic systems. 

The formalism of the transverse field method is simple to implement in any code that handles non-collinear magnetism. It can be used with or without the generalised Bloch theorem. The advantage of using the spin-spiral formalism is the possibility of calculating interactions between pairs of atoms that are not both contained in the unit cell. The method has several clear advantages over the conventional frozen magnon method besides the improved scaling and avoidance of the LWA. It is reasonable to expect the method to produce more accurate results than non-self-consistent approaches for systems where inter-atomic non-collinearity results in sizeable intra-atomic non-collinearity. Furthermore, more accurate results can be expected for systems with large induced moments, that depend sensitively on the directions of the surrounding moments. Simple rotations of the potentials at the sites of the magnetic sub-lattices are in these cases not likely to capture the intricate changes of the self-consistent potentials and corresponding total energy differences. 

When the Heisenberg model is expanded and applied to supercell calculations it is clear that higher order interactions play a crucial role in an accurate description of the non-collinear total energy landscape of bcc Fe. A significant increase in accuracy is obtained already when only bi-quadratic interactions are added to the regular Heisenberg model. The promising results obtained by the full multi-spin model indicates that the ambition to model the magnetic phase space for itinerant magnetic systems in a single model is reasonable. The advantages of the transverse field method is evident in the scaling and the possibility of evaluating and optimising the predictive power of the derived model against self-consistently derived quantities using the cross-validation analysis. The present study supplies a fully automatic, affordable and self-consistent method to systematically investigate exchange interactions beyond the standard Heisenberg description. This approach lends itself naturally to massive hightroughput calculations to find new useful or exotic magnetic materials.    

The perfect match between the self-consistent results for the regular Heisenberg Hamiltonian and the DLM results for bcc Fe is a subject for further studies. It would be interesting to see how general this phenomenon is. For materials where the longitudinal fluctuations are strong this equality probably breaks down. A natural continuation of the present study is the formulation of a general framework that makes use of the transverse-field metod and includes longitudinal fluctuations as well. In this way, one can properly describe the magnetic properties of real materials including temperature effects.  
 

\section*{Acknowledgements}

The authors wish to thank Prof. Olle Eriksson, Dr. Yaroslav Kvashnin, Dr. Lars Nordstr\"{o}m, Dr. Phivos Mavropoulos, Dr. Gustav Bihlmayer and Dr. Jonas Jacobsson for valuable discussions. This work was partly supported by the Kempe Foundation (SMK-1430, JCK-1605) and Knut and Alice Wallenberg Foundation, Sweden. We gratefully acknowledge the support of J\"ulich Supercomputing Centre in allocating resources to project is JIFF3800 and the National Supercomputer Centre in Sweden for the allocation of computing time for SNIC 2015/16-30 and SNIC 2016/1-44 .


\end{document}